\documentclass[lettersize,journal]{IEEEtran}
\usepackage{amsmath,amsfonts}
\usepackage{array}
\usepackage[caption=false,font=normalsize,labelfont=sf,textfont=sf]{subfig}
\usepackage{textcomp}
\usepackage{stfloats}
\usepackage{url}
\usepackage{verbatim}
\usepackage{graphicx}
\usepackage{cite}
\usepackage{bm}
\usepackage{varioref}
\usepackage{graphicx}
\usepackage{courier}
\usepackage{color}
\usepackage{xcolor}
\usepackage{colortbl}
\usepackage{multirow,hhline}
\usepackage{booktabs}
\usepackage[figuresright]{rotating}
\usepackage{makecell,rotating}

\usepackage{extarrows}
\usepackage{bbding}

\usepackage{multirow}

\usepackage{hyperref}
\usepackage{amsthm}
\usepackage{subfig}

\usepackage[ruled, linesnumbered]{algorithm2e}

\usepackage{algorithm}
\usepackage{algorithmic}

\usepackage{amssymb}

\usepackage{tabularx} 
\usepackage{ragged2e} 
\usepackage{booktabs} 

\usepackage{tikz}

\definecolor{Gainsboro}{RGB}{220,220,220}
\definecolor{FloralWhite}{RGB}{255,250,240}
\definecolor{Honeydew}{RGB}{240,255,240}
\definecolor{Lavender}{RGB}{230,230,250}

\begin{document}

\title{Efficient Bipartite Graph Embedding Induced by Clustering Constraints}

\author{Shanfan Zhang, Yongyi Lin, Yuan Rao, Zhan Bu
\thanks{\emph{First Author and Second Author contribute equally to this work.}}
\thanks{\emph{Corresponding author: Yuan Rao.}}
\thanks{Shanfan Zhang and Yuan Rao are with the School of software Engineering, Xi’an Jiaotong University, Xi’an, China (e-mail: zhangsfxajtu@gmail.com).}
\thanks{Yongyi Lin is with the Department of Mathematics, Xi’an Jiaotong University, Xi’an, China (e-mail: linyongyi@stu.xjtu.edu.cn).}
\thanks{Zhan Bu is with the School of Computer Science, Nanjing Audit University, Nanjing, China (e-mail: zhanbu@nau.edu.cn).
}

}

\markboth{Journal of \LaTeX\ Class Files,~Vol.~14, No.~8, August~2021}%
{Shell \MakeLowercase{\textit{et al.}}: A Sample Article Using IEEEtran.cls for IEEE Journals}

\IEEEpubid{0000--0000/00\$00.00~\copyright~2021 IEEE}

\maketitle

\begin{abstract}
Bipartite graph embedding (\emph{BGE}) maps nodes to compressed embedding vectors that can reflect the hidden topological features of the network, and learning high-quality \emph{BGE} is crucial for facilitating downstream applications such as recommender systems. However, most existing methods either struggle to efficiently learn embeddings suitable for users and items with fewer interactions, or exhibit poor scalability to handle large-scale networks. In this paper, we propose a \emph{C}lustering \emph{C}onstraints induced \emph{BI}partite graph \emph{E}mbedding (\emph{CCBIE}) as an integrated solution to both problems. \emph{CCBIE} facilitates automatic and dynamic soft clustering of items in a top-down manner, and capturing macro-preference information of users through clusters. Specifically, by leveraging the cluster embedding matrix of items, \emph{CCBIE} calculates the cluster assignment matrix for items and also captures the extent of user preferences across different clusters, thereby elucidating the similarity between users and items on a macro-scale level. \emph{CCBIE} effectively preserves the global properties of bipartite graphs, maintaining the cluster structure of isomorphic nodes and accounting for long-range dependencies among heterogeneous nodes. Our approach significantly enhances user-item collaborative relation modeling by integrating adaptive clustering for relationship learning, thereby markedly improving prediction performance, particularly benefiting cold users and items. Extensive experiments indicate that \emph{CCBIE} consistently and significantly improves accuracy over baseline models, particularly on sparse graphs, while also enhancing training speed and reducing memory requirements on large-scale bipartite graphs.
\end{abstract}


\section{Introduction}
Bipartite graph is a data structure that arises from modeling relationships between two types of entities (nodes). Interactions between nodes are restricted to occur solely between nodes of different types, with no direct connections between nodes of the same type. In recent years, Bipartite Network Embedding (\emph{BGE})~\cite{TransGNN,STERLING,AdaGCL} has garnered significant attention from both academia and industry due to its versatile applicability in various scenarios. Examples include image object recognition~\cite{image}, product recommendation for online shopping platforms~\cite{CEL}, and more. High-quality \emph{BGE} can greatly enhance the accuracy of downstream machine learning tasks, e.g., top-N recommendation~\cite{LightGCL,DR-GNN} and link prediction~\cite{GEBE,AnchorGNN}, and potentially yielding substantial economic returns and driving technological advancements.

\IEEEpubidadjcol

Researchers have proposed two main approaches to enhance the embedding quality of \emph{BGE}: (1) random walk-based methods~\cite{BINE,GEBE}, which rely on designing random-walk heuristics to construct approximate path contexts. (2) reconstruction-based methods~\cite{LightGCN,AnchorGNN}, akin to collaborative filtering models, aim to employ various Graph Neural Network (\emph{GNN}) models for reconstructing user-item interactions. While the above approaches have achieved some progress in terms of embedding quality, they primarily concentrate on learning the local graph structure. This involves assuming that nodes within proximity paths or neighborhoods are closely related. However, these methods often lack the capability to effectively model the global characteristics of bipartite graphs. This includes capturing the structure of clusters of isomorphic nodes and addressing the long-range dependencies of heterogeneous nodes. An associated issue stemming from this limitation is that for cold users and items with inadequate historical interactions, these models struggle to learn embeddings that generalize effectively. The challenges posed by data sparsity and skewed distributions significantly hinder their ability to model accurate user-item interactions. Consequently, this results in a pronounced decline in predictive performance as the scarcity of user data becomes evident.

\begin{figure}
    \centering
    \includegraphics[width=0.99\linewidth]{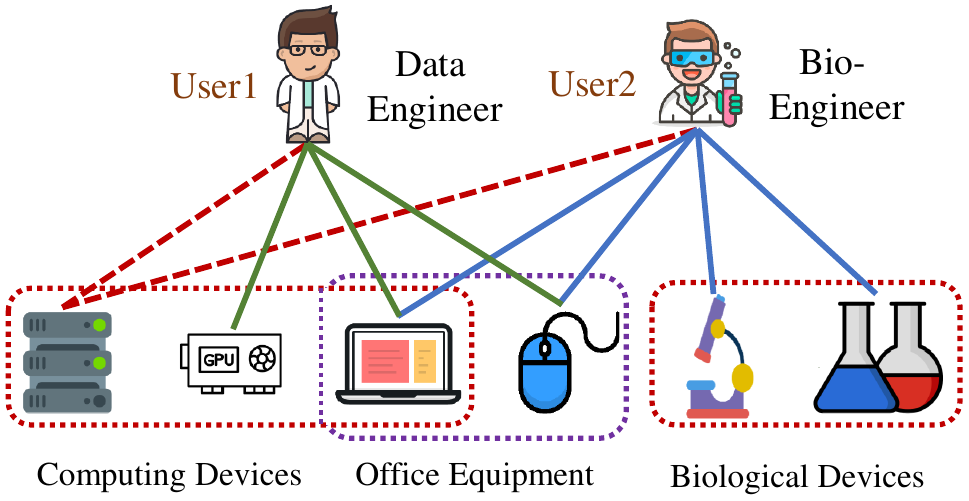}
    \caption{An illustrative example of user-item bipartite graph, along with its distinctive local and global properties, is provided. Locally, solid lines represent explicit inter-type connections, which are derived from historical interaction data. On a global scale, the dotted boxes denote potential interconnections among items, reflecting the community structure wherein items may exhibit shared genres. Additionally, the red dashed lines signify the long-term dependency between users and items.}
    \label{FIG:EXAMPLE}
\end{figure}

Cluster-level semantics serve as a valuable tool for enhancing the representation of diverse user preferences, thereby leading to improvements in prediction accuracy. As depicted in Fig.\ref{FIG:EXAMPLE}, by analyzing users' historical interaction data, one can infer their preference intensities across various product categories—such as computing devices, office equipment, and bio-devices—thus enabling the refinement of recommendation strategies. For example, a user who purchases a GPU is likely to exhibit a preference for related high-performance computing devices, such as servers. Similarly, users who purchase microscopes are more inclined to seek out reagent vials rather than servers. Consequently, in the recommendation process, the system should increase the prominence of computing devices for user 1 while diminishing the focus on bio-devices, and the reverse should apply for user 2.

To construct a \emph{BGE} solution capable of accurately modeling the recommendation process described above, it is imperative to address the following critical challenges:

\textbf{Challenge \uppercase\expandafter{\romannumeral1}: \emph{How to dynamic and automatic clustering under resource constraints?}} This stage of the operation involves identifying potential interconnections among homogeneous nodes, analogous to the dotted boxes shown in Fig.\ref{FIG:EXAMPLE}, which represent community structures wherein items may exhibit shared genres. To accomplish this, regularization techniques are often utilized, such as constraining all acquired embeddings to conform to a clustering structure. However, prevalent models~\cite{cluster-fixed,cluster-fixed-2} in current use often depend on predetermined clustering structure, which may detrimentally impact their performance and restrict adaptability. Moreover, these approaches necessitate full embedding training, leading to significant memory and time expenses.

\textbf{Challenge \uppercase\expandafter{\romannumeral2}: \emph{How to integration of cross-category cluster information?}} As depicted in Fig.\ref{FIG:EXAMPLE}, users demonstrate divergent purchase preferences for different categories of items (e.g., electronic versus biological devices) influenced by their professional specializations (e.g., data engineers versus bio-engineers). By analyzing these preferences at the cluster level, informed by historical interaction data, the system can deliver more nuanced, personalized recommendations that align with the distinct requirements of various user. There remains a deficiency in mechanisms for effectively facilitating the desirable transfer of cluster information across categories, leading to computational inefficiencies and potential semantic inaccuracies. For instance, STERLING~\cite{STERLING} aims to capture cluster-level synergies between different node types by maximizing mutual information between user clusters and item clusters. Nevertheless, this approach falls short in transmitting cluster-level semantics to individual users, which significantly degrades the model's predictive performance and exacerbates its complexity.


\textbf{Challenge \uppercase\expandafter{\romannumeral3}: \emph{How to efficiently combine explicit and implicit relationships between users and items to achieve more accurate relationship modeling?}} In essence, prevailing \emph{BGE} models~\cite{mainstream-1,mainstream-2,LightGCN,AnchorGNN} endeavor to encapsulate both explicit relationships and implicit synergies between nodes through a singular relational framework. This approach is fraught with challenges and appears inadequate, as it may fail to fully encompass the intricate and multifaceted nature of node interactions. One notable pitfall of this computational paradigm is the disparity between the distribution of nodes in the embedding space and their distribution within the actual network. This inconsistency can impede the ability of the embedding vectors to effectively capture the contextual information of the nodes. For instance, while the embedding space allows for connections only between nodes with a sufficiently high similarity, real networks often establish links between multi-hop neighbors due to the presence of higher-order structures, such as community structures.

In pursuit of a more flexible, highly scalable, and superior-performing \emph{BGE} solution, while addressing the aforementioned challenges, this paper introduces a novel \emph{C}lustering \emph{C}onstraints induced \emph{BI}partite graph \emph{E}mbedding (\emph{CCBIE}) model. This model is specifically designed to capture both local and global properties of bipartite graphs with improved efficacy. For the microstructure, akin to other models, \emph{CCBIE} maximizes the similarity of positive node embedding pairs. In the context of macrostructure, \emph{CCBIE} utilizes an end-to-end deep clustering model to generate clustering assignment matrices for both node types. Subsequently, global cluster synergies are captured through a simple yet effective integration scheme. Specifically, \emph{CCBIE} delineates the following three operations to tackle each of these challenges:

\textbf{Solution \uppercase\expandafter{\romannumeral1}:} \emph{CCBIE} introduces a cluster embedding matrix, eliminating the need for full embedding training to determine each item's affinity to specific clusters. Through end-to-end training, \emph{CCBIE} updates relevant embeddings via back-propagation during training losses. Once cluster and item embeddings are optimized, cluster allocations are recalculated to better reflect entity interactions, achieving fine-grained automatic cluster optimization. In \emph{CCBIE}, all entities can utilize and collectively learn shared embeddings, thereby significantly reducing memory costs.

\textbf{Solution \uppercase\expandafter{\romannumeral2}:} Given \emph{CCBIE}'s aim to capture user preferences across various categories of items, utilizing the cluster embedding matrix—which encapsulates the distribution of different cluster—is a logical approach. A natural indicator for user preference would be the similarity between the user embedding and the cluster embedding associated with the respective category. Immediately following, the implicit user preference for a specific item should be influenced by two factors: the variation in the user's preference across different categories of items, and the likelihood that the item falls into each of those categories. To harness global cluster synergies, \emph{CCBIE} proposes an efficient and validated approach to integrating cross-category information. This method enables continuous optimization of user category preferences through error back-propagation during the training process.

\textbf{Solution \uppercase\expandafter{\romannumeral3}:} Compared to existing models, \emph{CCBIE} alleviates the embedding restriction between node pairs by not enforcing that the similarity (explicit relation) of positive node pairs exceeds that of negative node pairs. For instance, although the embedding similarity between negative node pairs may surpass that of positive node pairs, \emph{CCBIE} can still deliver high-quality prediction performance if the implicit relationship—reflected by global clustering synergy—between negative node pairs is substantially weaker.

\emph{CCBIE} excels at modeling global synergies within bipartite graphs, all while ensuring robust scalability. Specifically, we rationalize that the overall time complexity and space complexity of \emph{CCBIE} are proportional to the number of edges and nodes in the network, respectively. Extensive experiments demonstrate that \emph{CCBIE} outperforms baselines in both link prediction and top-$N$ recommendation tasks, particularly on sparser networks, where \emph{CCBIE} outperforms the best competitor by up to 10.35\% in accuracy. 

\section{Related Work}

A straightforward approach to address the \emph{BGE} problem is to treat the bipartite graphs as typeless homogeneous graphs and then utilize established solutions, such as \emph{LINE}~\cite{LINE}, to derive the embedding outcomes. However, such solutions are not tailored for bipartite graphs, and thus disregarding the distinctive structural characteristics and type-specific information, consequently yielding low-quality \emph{BGE} results.

In the pursuit of higher quality embeddings, an effective line of research involves generating node embeddings through the establishment of similarity and proximity metrics over the constructed approximation paths. For instance, \emph{BiNE}~\cite{BINE} maintains the long-tailed distribution of nodes by performing numerous biased random walks and generates node embeddings by incorporating both edges and paths in the input graph. \emph{GEBEp}~\cite{GEBE} learns to maintain multi-hop similarity and proximity by assigning importance to specific paths using a probabilistic mass function, and then optimizes the objective function through eigen-decomposition.

Another promising approach involves leveraging deep learning techniques such as graph convolutional neural networks, self-augmented learning and transformers to enhance the quality of \emph{BGE}. Specifcally, \emph{NGCF}~\cite{NGCF} incorporates user-item interactions within the message propagation process to capture collaborative signals and higher-order connectivity. \emph{LightGCN}~\cite{LightGCN} is an efficient and lightweight model focused exclusively on neighborhood aggregation, given that extensive ablation experiments have validated that the operations of feature transformation and nonlinear activation, do not contribute to, and may even degrade the effectiveness of \emph{NGCF}.

\emph{NGCF}~\cite{NGCF} and \emph{LightGCN}~\cite{LightGCN} are typically designed to rely solely on observing interaction labels for model training, primarily focusing on modeling local graph structures in the latent space. However, this approach hinders their ability to effectively handle sparse interaction graphs or those with significant global properties. Recently, researchers have initiated exploration into employing various self-supervised learning techniques to enhance the robustness or generalization capabilities of the model. For example, \emph{BiGI}~\cite{BiGI} enhances the relevance of learned node embeddings on a global scale by maximizing the mutual information between the global representation, composed of two prototypical representations, and the local representation encoded from sampled edges. \emph{SHT}~\cite{SHT} integrates hypergraph neural networks with topology-aware transformers to capture global cross-user collaborative relationships. \emph{HCCF}~\cite{HCCF} develops a hypergraph-enhanced cross-view contrastive learning framework that captures the intrinsic relationships between users and projects through iterative local neighborhood aggregation and global message propagation. \emph{AdaGCL}~\cite{AdaGCL} utilizes two adaptive contrastive view generators for data augmentation, aiming to rectify the challenges posed by noisy and skewed user behavior data in real-world recommendation scenarios, thereby boosting the collaborative filtering paradigm. \emph{LightGCL}~\cite{LightGCL} explores the concept of leveraging singular value decomposition to enhance user-item interaction graph structures, enabling unconstrained structural refinement for modeling global synergistic relationship and demonstrating a strong capability to counteract data sparsity and popularity bias. TransGNN~\cite{TransGNN} employs transformer layers to expand the receptive field and decouple message aggregation from links, which enables the aggregation of messages from more relevant nodes, thus optimizing the message-passing process of \emph{GNNs}.

\section{Preliminaries}

\textbf{Symbol description.} A bipartite graph can be represented as a triple $\mathcal{A} = \left ( \mathcal{U},\mathcal{V},\textbf{E} \right )$, where $\mathcal{U}= \left \{ u_{1},u_{2},\dots ,u_{I}  \right \}$ denotes the set of $I$ users (source nodes), $\mathcal{V}= \left \{ v_{1},v_{2},\dots ,v_{J}  \right \}$ denotes the set of $J$ items (destination nodes), $\textbf{E} \subseteq \mathbb{R} ^{I\times J}$ denotes the observed interactions between users and items, such that $\textbf{E}_{ij} = 1$ if $u_{i}$ has adopted item $v_{j}$, otherwise $\textbf{E}_{ij} = 0$. We use bold uppercase letters for matrices (e.g., \textbf{U}) and bold lowercase letters for vectors (e.g., \textbf{u}). 

\textbf{Problem definition.} Based on the above definition, a GNN-based \emph{BGE} model can be formalized as an inference model: (1) \textbf{Inputs} user-item interaction data $\textbf{E}$; (2) \textbf{Models} the interaction patterns between pairs of entities based on the observed historical interactions $\textbf{E}^{O}$ and \textbf{maps} each node $u_{i} \in \mathcal{U} $ and $v_{j} \in \mathcal{V} $ to informative $d$-dimensional vector representations $\textbf{u}_{i} \in \mathbb{R} ^{d}$ and $\textbf{v}_{j} \in \mathbb{R} ^{d}$, respectively; (3) \textbf{Outputs} predictions for unobserved interactions between user-item pairs $\textbf{E}^{U}$ ($\textbf{E}= \textbf{E}^{O}\cup \textbf{E}^{U}$ and $\textbf{E}^{O}\cap \textbf{E}^{U}= \varnothing$) using the learned embeddings.

\section{Methodology}

In this section, we first embed local structural information into potential node representations and outline an approach to explore cross-category global synergies using a novel clustering-based structural capture strategy, then introduce a loss function incorporating cluster-prototype disentanglement to facilitate end-to-end training, and finally present a rational analysis of the proposed methodology.

\begin{figure*}
\center
\linespread{0.1}
\includegraphics[width=0.99\textwidth]{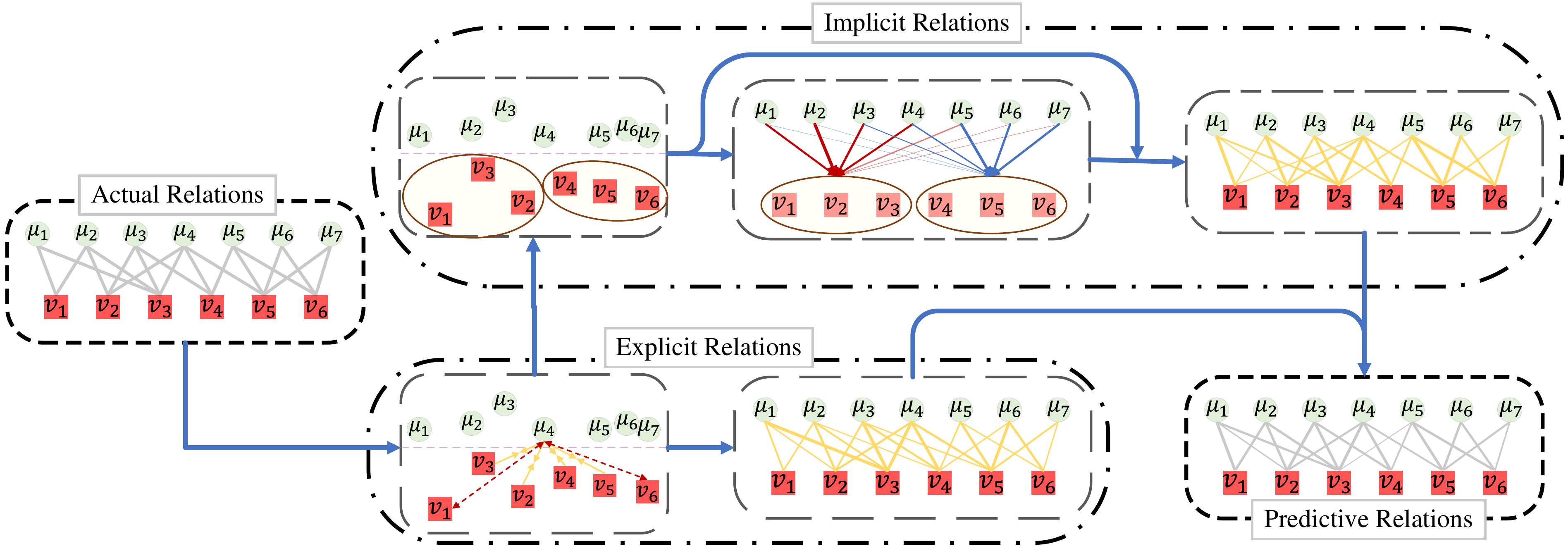}
\caption{Overall framework of the proposed \emph{CCBIE} model. The orange-shaded area enclosed by the elliptical box represents a underlying cluster structure. The thickness of the edges reflects the correlation score between nodes. A thicker edge suggests a stronger correlation and a higher likelihood of link formation between the nodes.}
\label{prediction}
\end{figure*}

\begin{figure*}
\center
\linespread{0.1}
\includegraphics[width=0.99\textwidth]{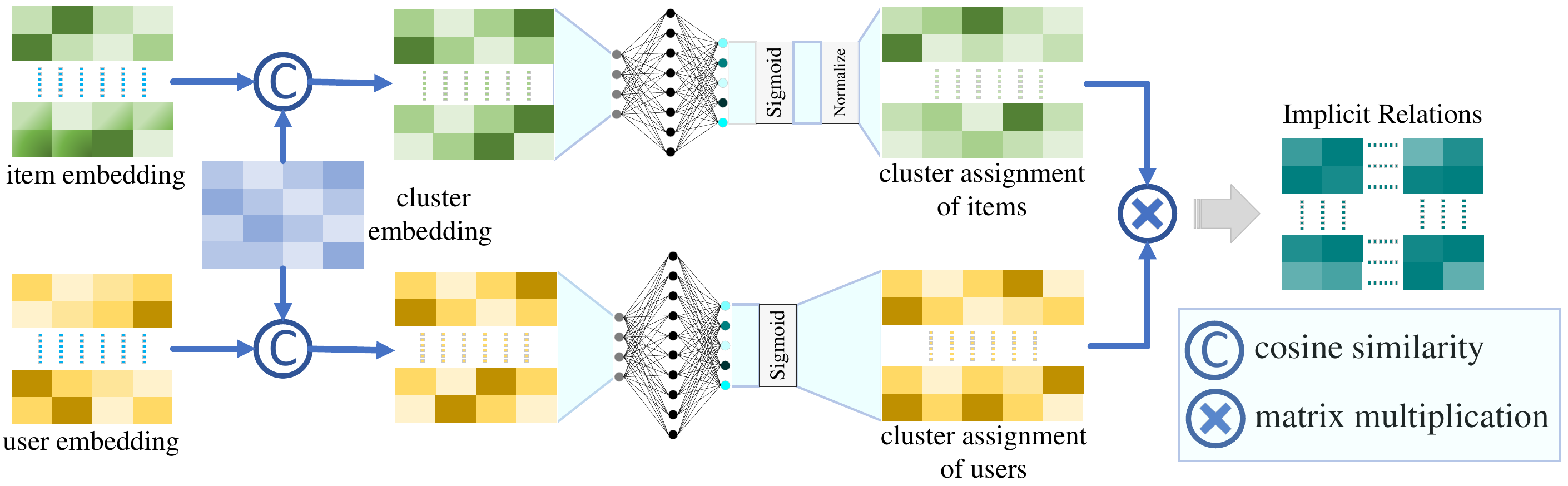}
\caption{Workflow of the proposed cluster-aware cross-category integration module. In addition to the user and item embedding matrix, we propose the incorporation of a cluster embedding matrix. This additional matrix is designed to capture and elucidate the nuanced connections between items and user preferences at the meso level. It can be updated through error backpropagation during subsequent training, thereby enhancing prediction accuracy and automating the clustering process.}
\label{cluster}
\end{figure*}

\subsection{Graph Structure Learning}

In this subsection, we elaborate on the proposed \emph{CCBIE} framework, which effectively harnesses both local and global synergies in bipartite graphs. \emph{CCBIE} incorporates local structural information into node representations and employs a cluster-aware cross-category integration module for global relationship extraction. Fig.\ref{prediction} illustrates the comprehensive workflow.

\subsubsection{Modeling Explicit Relations}

In bipartite graph networks, edges linking nodes of different types serve as explicit indicators for establishing relationships between nodes. We model these relationships explicitly by evaluating the local proximity between connected vertices. The objective is to minimize the similarity between negative pairs (e.g., randomly sampled node pairs) and maximize the similarity between positive pairs in the embedding space.

Cosine similarity serves as a widely utilized metric for comparing text similarities in applications like search engines and recommender systems. We follow this setting, and leverage cosine similarity between embedded vectors to model interactions between entities. Furthermore, we employ a normalization function to convert these interaction values into the probability space:
\begin{equation}
  \widetilde{\textbf{Y}}_{ij}= \left ( \frac{u_{i} \cdot v_{j}}{\left \| u_{i} \right \| \left \| v_{j} \right \| } + 1 \right ) /2 \in \left [ 0,1 \right ]  
\end{equation}
Essentially, minimizing the difference between the true distribution of inter-vertex co-occurrences and the reconstructed distribution ensures that vertices strongly connected in the original network remain closely positioned in the embedding space, thereby preserving their desired local proximity.

\subsubsection{Modeling Implicit Relations}

\emph{CCBIE} facilitates dynamic and automatic clustering of users and items, thereby potentially capturing higher-order structural information and addressing issues related to cold users and items. Within \emph{CCBIE}, all entities can utilize and collectively learn a shared embedding. Unlike \emph{CEL}~\cite{CEL}, \emph{CCBIE} simultaneously performs both embedding optimization and cluster optimization, significantly reducing the risk of converging to a locally optimal solution. This approach enhances convergence speed and yields superior performance results. Fig.\ref{cluster} summarizes the workflow of our cluster-aware cross-category integration module.

We denote matrices $\textbf{U} \in \mathbb{R} ^{I\times d}$ and $\textbf{V} \in \mathbb{R} ^{J \times d}$ as embedding vectors for all nodes in $\mathcal{U}$ and $\mathcal{V}$, respectively. To initiate item clustering, we introduce a cluster embedding matrix $\textbf{C} \in \mathbb{R} ^{K\times d} $, where each row represents an embedding at the center of a cluster. Subsequently, we define the affinity of an item to each cluster as the similarity between its embedding vector and the rows of matrix $\textbf{C}$. For instance, the propensity of item $v_{j}$ towards cluster ${k}$ can be computed as
	\begin{equation}
		\textbf{T}_{jk}
		=\frac{\textbf{v}_{j}\cdot \textbf{C}_{k}}{\left \| \textbf{v}_{j} \right \| \cdot \left \| \textbf{C}_{k} \right \|}
		=\frac{ {\textstyle \sum_{l=1}^{d}}\textbf{v}_{jl}\times \textbf{C}_{kl}}{\sqrt{ {\textstyle \sum_{l=1}^{d}}\left ( \textbf{v}_{jl} \right )^{2}} \times\sqrt{{\textstyle \sum_{l=1}^{d}}\left(\textbf{C}_{kl}\right)^{2}}}
	\end{equation}

Similar to \emph{AnchorGNN} \cite{AnchorGNN}, this operation can be interpreted as a node sending messages to all cluster centers using a parameterized function. It constructs virtual edges between the node and each cluster center, which depict potential spatial projections from the graph nodes to the cluster structure. Based on this calculation, we can determine the cluster propensity for all items and generate a clustering embedding matrix $\textbf{T} \in \mathbb{R} ^{J\times K}$.

To enhance the robustness and expressiveness of such virtual edges, multilayer perceptron (\emph{MLP}) is employed to further model the potential relationship between item nodes and cluster centers: $\textbf{T}^{M} = MLP\left ( \textbf{T} \right )  \in \mathbb{R} ^{J\times K_{q}}$. Note that we permit MLP to adjust the number of clusters within the graph as required, i.e., $K_{q} \ne K$.

The propensity of a node towards a specific cluster should lie within the interval of 0 to 1. Thus, the output values from the \emph{MLP} are transformed into the probability space by scaling them using sigmoid function:
	\begin{equation}
		\textbf{T}^{O}_{ij} = \frac{1}{1+ exp \left ( - \textbf{T}^{M}_{ij} \right )  }
	\end{equation}

As depicted in Fig.\ref{FIG:EXAMPLE}, for a given item, it does not exclusively belong to a single cluster. Rather, it can represent a hybrid combination of several clusters, each with varying weights. For instance, a computer may serve as both an office device and a high-performance computing tool. Therefore, we employ a soft cluster allocation approach where each element of $\textbf{T}^{O}$ ranges between 0 and 1, ensuring that the sum of elements in each row equals 1. Our primary goal is to achieve compact representations of nodes within individual clusters while optimizing the distinctiveness between nodes across different clusters.

We eschew the frequently employed softmax function for determining the cluster affiliation of nodes due to its tendency to hard-allocate the cluster distribution of nodes, leading to the loss of valuable structural information. In our experiments, we observe that effective performance in general can be achieved by utilizing the uniform normalization approach outlined in Eq.\ref{uni-max} for nodes' cluster propensity. Thus we refrain from incorporating specialized components aimed at optimizing cluster assignment, thereby avoiding unnecessary complexity in \emph{CCBIE} while preserving its simplicity.
\begin{equation}
    \label{uni-max}
    \textbf{T}^{V}_{ij} = \frac{\textbf{T}^{O}_{ij}}{ {\textstyle \sum_{l=1}^{K_{q}}} \textbf{T}^{O}_{il} }
\end{equation}

The next phase of \emph{CCBIE} aims to ascertain user preferences across different categories of items. From the preceding definition, the categorization of a item is established by evaluating the similarity or distance between its embedding vector and the cluster centers. Consequently, an intuitive approach involves quantifying this preference through the similarity or distance between the user's embedding vector and the cluster centers:
	\begin{equation}
		\textbf{P}_{ik}
		=\frac{\textbf{u}_{i}\cdot \textbf{C}_{k}}{\left \| \textbf{u}_{i} \right \| \cdot \left \| \textbf{C}_{k} \right \|}
		=\frac{ {\textstyle \sum_{l=1}^{d}}\textbf{u}_{il}\times \textbf{C}_{kl}}{\sqrt{ {\textstyle \sum_{l=1}^{d}}\left ( \textbf{u}_{il} \right )^{2}} \times\sqrt{{\textstyle \sum_{l=1}^{d}}\left(\textbf{C}_{kl}\right)^{2}}}
	\end{equation}
 
We opted to utilize \emph{MLP} to further augment the expressiveness of this preference, subsequently, we normalized its output through the sigmoid function:
	\begin{equation}
        \textbf{P}^{M}_{ij} = MLP\left ( \textbf{P} \right )  \in \mathbb{R} ^{I \times K_{q}}, \quad
	    \textbf{P}^{O}_{ij} = \frac{1}{1+ exp \left ( - \textbf{P}^{M}_{ij} \right )  }
	\end{equation}

We emphasize that, unlike items, a single user may have strong preferences for items across multiple clusters. Therefore, there is no necessity to soft-assign users' cluster preferences. Our subsequent ablation experiments further confirm that incorporating this operation not only diminishes the scalability of the model but also adversely affects its performance.

After obtaining the cluster allocation matrix for items ($\textbf{T}^{V} \in \mathbb{R} ^{J \times K_{q}}$) and the cluster preference matrix for users ($\textbf{P}^{O} \in \mathbb{R} ^{I \times K_{q}}$), \emph{CCBIE} integrates these matrices to capture higher-order structural patterns across node types, thereby enforcing cluster constraints on relationships between node pairs. The effectiveness and prevalence of \emph{transformers}~\cite{transformer} have inspired numerous studies to incorporate cross-attention layers for modeling interactions among entities belonging to different types. However, through extensive experimentation, we have determined that the quality of capturing global cluster synergies can be significantly enhanced by plainly employing weighted sums to amalgamate cluster information across nodes. Formally, the implicit relationship between user-item pairs is defined as the product of the cluster assignment matrix $\textbf{T}^{V}$ and the cluster preference matrix $\textbf{P}^{O}$:
	\begin{equation}
        \widehat{\textbf{Y} } =  \textbf{P}^{O} \textbf{T}^{V} \in \mathbb{R} ^{I \times J}
	\end{equation}
By definition, $\forall u_{i}\in \mathcal{U}$, $\textbf{P}^{O} _{i \cdot}\in \left [ 0,1 \right ]$; $\forall v_{j}\in \mathcal{V}$, $\textbf{T}^{V} _{j\cdot}\in \left [ 0,1 \right ]$ and ${\textstyle \sum_{l=1}^{K_{q}}} \textbf{T}^{V} _{jl} = 1$. Consequently, each element of $\widehat{\textbf{Y} }$ will fall within the interval from 0 to 1. Moreover, a higher value indicates greater suitability for establishing edges between pairs of entities.

\subsection{Model Training}

Upon our \emph{BGE} model incorporating clustering constraints, the relations between entities is decomposed into explicit and implicit components: the one-hop structural constraints between user-item pairs (explicit relations) is ensured by the node embedding similarities; each user and item node accesses global information via the clustering centers, thereby generating implicit higher-order relations. Thus, predicting the actual relation between entities becomes straightforward through the weighted sum of these two components:
	\begin{equation}
        \textbf{Y}= \alpha \widetilde{\textbf{Y}}  +  \left ( 1- \alpha  \right ) \widehat{\textbf{Y}}
        \label{relations}
	\end{equation}
\noindent where $\alpha$ is the hyperparameter. Any element $\textbf{Y}_{ij}$ in matrix $\textbf{Y}$ represents the probability of forming an edge between nodes $u_{i}$ and $v_{j}$ when given node $u_{i}$. Structure learning on bipartite graphs can be formulated as a maximum likelihood optimization problem: 
	\begin{equation}
        \max  {\textstyle \prod_{u_{i} \in \mathcal{U} }}  Pr\left ( \mathcal{N}\left ( u_{i} \right )\mid u_{i}   \right )
	\end{equation}
\noindent where $ Pr\left ( \mathcal{N}\left ( u_{i} \right )\mid u_{i}   \right ) =  {\textstyle \prod_{v_{j}\in \mathcal{N}\left ( u_{i} \right ) }} \textbf{Y}_{ij} $, $\mathcal{N} \left ( u_{i} \right )$ denotes the set of neighbors of node $u_{i}$ obtained by positive or negative sampling. Therefore, we can deduce the structure directly from the set of edges without needing to construct the adjacency matrix. We then construct the mean squared loss to maximize the probabilities for structure learning:
	\begin{equation}
        \mathcal{L}_{structure} =  {\textstyle \sum_{\left ( u_{i}, v_{j} \right ) \in \textbf{E}^{O} }} \left ( \textbf{E}_{ij} - \textbf{Y}_{ij}  \right )^{2}
	\end{equation}
Inspired by the pursuit of distinct and discriminative cluster centers, we propose the cluster disentanglement regularization, which draws upon insights from the advances of disentangled representation learning~\cite{CLUSTER-LOSS,CLUSTER-LOSS-2}:
	\begin{equation}
        \mathcal{L}_{cluster} =  - \beta  {\textstyle \sum_{i=1}^{K- 1}} {\textstyle \sum_{j=i+  1}^{K}} \left \| \textbf{C}_{i}- \textbf{C}_{j}  \right \| ^{2}
        \label{cluster-sep}
	\end{equation}
\noindent where $\beta$ is also the hyperparameter. Therefore, the final optimization objective of \emph{CCBIE} is:
	\begin{equation}
        \mathcal{L}_{obj} = \mathcal{L}_{structure} + \mathcal{L}_{cluster}
	\end{equation}

\subsection{Model Analysis}

The proposed \emph{CCBIE} model can be summarized as the following two-stage schema:
\begin{equation}
  \textbf{u}_{i}, \textbf{v}_{j}, \textbf{C} \gets Embed\left ( u_{i},v_{j} \mid \textbf{E}^{O} \right )  
\end{equation}
\begin{equation}
  \textbf{Y}_{ij}\gets Predict\left ( \textbf{u}_{i}, \textbf{v}_{j}, \textbf{C} \right ) 
\end{equation}
Stage $Embed\left ( \cdot \right )$ involves the embedding process where user $u_{i}$ and item $v_{j}$ are projected into a $d$-dimensional hidden space using historical interaction information. Simultaneously, cluster-prototype embeddings $\textbf{C} \in \mathbb{R} ^{K\times d} $ are generated to capture global structural information. Here $K$ denotes the number of clusters. Stage $Predict\left ( \cdot \right )$ aims to utilize the learned embedded data $\textbf{u}_{i}$, $\textbf{v}_{j}$, and $\textbf{C}$ to predict the user-item relations using the prediction score $\textbf{Y}_{ij}$.

The node-cluster \textbf{\emph{m}}essage \textbf{\emph{p}}ropagation paradigm (\textbf{\emph{MP}}) introduced by \emph{CCBIE} circumvents the reliance on multiple neighborhood typical in \emph{GCN}-based models, thereby enhancing its capability to process large-scale graphs. Throughout training, \emph{CCBIE} maintains global context by continuously learning and optimizing each cluster center based on the input graph. Correspondingly, \emph{CCBIE} effectively utilizes global information to facilitate the generation of distinct structural features.


\textbf{Time complexity.} \emph{CCBIE}’s time complexity is mainly related to that of $\mathcal{L}_{structure}$, $\mathcal{L}_{cluster}$ and node-cluster \textbf{\emph{MP}}. On general graphs trained in a mini-batch manner, the time complexity of $\mathcal{L}_{structure}$ is $\mathcal{O} \left ( B \left ( N +1  \right ) \right )$, where $B$ denotes batch size, $N$ functions as a hyperparameter designed to regulate the magnitude of the negative sample set (each positive sample in $\textbf{E}^{O}$  corresponds to $N$ negative samples). The computation time required for evaluating $\mathcal{L}_{cluster}$ is $\mathcal{O} \left ( K^{2}d  \right )$. Assuming the architecture of the \emph{MLP} used for both users and items is $K\times K_{h} \times K_{q}$, $d \ge K_{h} \ge  K \ge  K_{q}$, the time complexity of node-cluster \textbf{\emph{MP}} is $\mathcal{O} \left ( 2B\left ( K\times d +  K\times K_{h} + K_{h}\times K_{q} + K_{q} \right ) \right )$, bounded by $\mathcal{O} \left ( 2BKd\right )$. Consequently, the overall time complexity is $\mathcal{O}\left ( 2\left | E \right |Kd  \right ) $ because, in subsequent experiments, we set $B$ to be greater than $K$.

\textbf{Space complexity.} The overall space complexity encompasses parameters for storing user and item nodes, alongside additional parameters within the \textbf{\emph{MP}}. Specifically, the overall space allocated is adequate $\mathcal{O} \left ( \left ( I+  J +  K\right ) \times d +  2\times \left ( K\times K_{h} + K_{h}\times K_{q} \right ) \right ) $, which can be reduced to $\mathcal{O} \left ( \left ( I+  J +  K\right ) \times d \right )$.

\begin{table}[H]
\caption{Statistics of Datasets}
\centering
\label{datasets}
\footnotesize
\renewcommand{\arraystretch}{1.0}
\begin{tabular}{||c||c||c||c||c||}

\hline

Dataset
& User \#
& Item \# 
& Interaction \#
& Density

\\

\hline

DBLP & 6,001 & 1,308 & 29,256  & $3.7\times 10^{-3}$   \\

ML-100K  & 943   & 1,682  & 100,000  & $6.3 \times 10^{-2}$  \\

ML-10M   & 69,878   & 10,677  & 10,000,054  & $1.3 \times 10^{-2}$  \\

\hline

Wikipedia & 15,000 & 3,214 & 64,095 & $1.3 \times 10^{-3}$ \\

YELP & 42,712 & 26,822 & 182,357 & $1.6 \times 10^{-4}$ \\

Pinterest & 55,187  & 9,916  &  1,500,809 & $2.7 \times 10^{-3}$ \\

\hline

\end{tabular}
\end{table}

\section{Evaluation}

\subsection{Setting and Implementation.} 

Table \ref{datasets} presents specific statistics for the six public datasets utilized in the experiments. We adopted the identical experimental setup utilized in \emph{BiGI} for \emph{DBLP}, \emph{ML-100K} and \emph{ML-10M}. Following the experimental setup established in prior research~\cite{BINE,GEBE}, for the link prediction task, we randomly sample 40\% of the edges from the input graph and an equivalent number of non-existent edges to form the test set.

We utilize 13 baseline methods, organized into three categories, to conduct a comparative analysis with \emph{CCBIE}. The categories are: (i) \emph{GNN}-based methods, including \emph{LightGCN}, \emph{BiGI}, \emph{SHT}, \emph{HCCF}, \emph{AdaGCL}, \emph{CEL-NMF}~\cite{CEL}, \emph{STERLING}, \emph{DR-GNN}~\cite{DR-GNN}, \emph{LightGCL} and \emph{TransGNN}; (ii) metric-based methods, specifically \emph{GEBEp}~\cite{GEBE} and \emph{BiNE}~\cite{BINE}; and (iii) a homogeneous network embedding method, \emph{LINE}~\cite{LINE}. For each of these models, we use the open-source implementation under default settings.

We employed the PyTorch framework to implement our model. We adopted the identical experimental setup utilized in \emph{BiGI}~\cite{BiGI} for \emph{DBLP}, \emph{ML-100K} and \emph{ML-10M}. Following the experimental setup established in prior research~\cite{BINE,GEBE}, for the link prediction task, we randomly sample 40\% of the edges from the input graph and an equivalent number of non-existent edges to form the test set. To facilitate fair comparisons, we standardized the embedding dimensions for all competitors to $d = 64$. The embedding vectors for the nodes were initialized using the Xavier method, and \emph{CCBIE} was trained utilizing the Adam optimizer with a learning rate set to 0.002. For all datasets, the batch size $B$ was set to 256. Additionally, we utilized a two-layer neural network to further reinforce the representations of $\textbf{T}^{M}$ and $\textbf{P}^{M}$, respectively. To optimize the training process, we implement a negative sampling strategy in accordance with the method described in \emph{AnchorGNN}~\cite{AnchorGNN}. For each source node $u_{i}$ within the training set $\textbf{E}^{O}$, we construct the negative sample set $\mathcal{N}_{\mathcal{S} }\left ( u_{i} \right )$ by uniformly sampling $N$ nodes from the entire destination node set $\mathcal{V}$, where $N$ is a hyperparameter. To ensure fairness, all experiments are conducted on the same Linux server with an Intel(R) Xeon(R) CPU E5-2680 v3 @ 2.50GHz CPU, 251G RAM and a GeForce RTX 4090 GPU (24GB). All source code, datasets, and other relevant artifacts from our experiments will be made publicly available on GitHub.


\begin{table*}[htbp]
    \centering
    \caption{ Top-$N$ recommendation performance (\%). \textcolor{red}{\textbf{Bold}} indicates the best result and \textcolor{blue}{\underline{underline}} indicates the second best. ($N$: $NDCG$, $R$: $RECALL$)}

    \renewcommand{\arraystretch}{1.3} 
    \setlength\tabcolsep{3pt} %
    \resizebox{\linewidth}{!}{
    \begin{tabular}{c|c|cccccccccccccc}
	\toprule
	
	\textbf{Datasets}
	&    \textbf{Metrics}   
	&    \emph{LINE}    &  \emph{BiNE}           &   \emph{GEBEp} 
	&    \emph{LightGCN}      &  \emph{BiGI}     &   \emph{SHT}     
    &    \emph{HCCF}          &   \emph{AdaGCL}      
    &  \emph{CEL-NMF}      &   \emph{STERLING}   &  \emph{DR-GNN}   &  \emph{LightGCL}
    &   \emph{TransGNN} 
    &    \emph{CCBIE}   \\ 
	
	\midrule
				
	\multirow{6}{*}{\emph{DBLP}}  
	&	$N@3$   
	& 16.88     & 19.85     & 24.07
	& 27.44    & 23.60     & 27.42
    & 25.36        & 21.13
    & 16.25     & 20.42   &  25.57   & 22.19   &  \textcolor{blue}{\underline{29.67}}
    & \textcolor{red}{\bm{$32.74$}}    \\
	
	\multirow{6}{*}{}  
	&	$N@5$   
	& 19.57     & 21.95     & 25.22
	& 28.41  &  25.28    & 27.52
    & 25.44         & 21.77
    & 19.80     & 21.90   & 25.55  & 22.94  &  \textcolor{blue}{\underline{31.72}}
    & \textcolor{red}{\bm{$34.00$}}    \\

	\multirow{6}{*}{}  
	&	$N@10$   
	& 20.46     & 25.15     & 26.95
	& 30.07    & 28.20     
    & 29.20
    & 28.63         & 23.51
    &  25.10    & 23.36    & 28.01  & 24.78  &  \textcolor{blue}{\underline{34.58}}
    & \textcolor{red}{\bm{$35.99$}}    \\
				
	\multirow{6}{*}{}  
	&	$R@3$   
	& 18.69     & 21.38     & 22.46
	& 24.91    & 23.95     & 24.78
    & 23.08         & 19.39
    & 14.27     & 21.07     & 21.98   & 20.52   &  \textcolor{red}{\bm{$30.86$}}
    & \textcolor{blue}{\underline{30.52}}    \\

    \multirow{6}{*}{}  
	&	$R@5$   
	& 22.29     & 25.85     & 26.19
	& 27.65    & 26.65     & 28.18
    & 26.64         & 23.37
    & 19.06     &  26.45    & 24.94  & 24.68    &  \textcolor{red}{\bm{$37.85$}}
    & \textcolor{blue}{\underline{37.37}}    \\

    \multirow{6}{*}{}  
	&	$R@10$   
	&  26.76    & 29.46     & 30.01
	&  34.61   & 32.43     & 33.49
    &  33.50       & 29.11
    & 28.14     & 31.03      & 32.11  &  30.90    &  \textcolor{red}{\bm{$45.76$}}
    & \textcolor{blue}{\underline{44.84}}    \\
				
	\midrule
		
	\multirow{6}{*}{\emph{ML-100K}}  
	&	$N@3$   
	& 7.42     &  7.51    & 52.17
	& \textcolor{blue}{\underline{60.69}}    &  12.28    & 58.91
    & 57.01         & 56.96
    &  13.02    & 13.89  & 53.38  & 55.13    &  53.67
    & \textcolor{red}{\bm{$62.74$}}    \\
	
	\multirow{6}{*}{}  
	&	$N@5$   
	& 10.37     &  9.66    & 48.22
	& \textcolor{blue}{\underline{56.69}}    &  15.76    & 55.96
    & 54.78         & 54.15
    & 15.61     & 17.96   & 51.46  & 53.66  &  52.65
    & \textcolor{red}{\bm{$59.54$}}    \\

	\multirow{6}{*}{}  
	&	$N@10$   
	& 16.26     & 13.59     & 42.64
	& \textcolor{blue}{\underline{52.96}}    & 22.06     & 51.94
    & 49.75         & 50.83
    &  20.56    & 24.05  & 48.71   & 50.54   &  48.77
    & \textcolor{red}{\bm{$54.88$}}    \\

	\multirow{6}{*}{}  
	&	$R@3$   
	& 1.85     & 1.82     & 6.11
	& \textcolor{blue}{\underline{7.48}}    & 3.02     & 7.23
    & 6.80         & 7.24
    & 1.70     &  3.26  & 5.95  & 7.01   &  6.58
    & \textcolor{red}{\bm{$7.49$}}    \\

    \multirow{6}{*}{}  
	&	$R@5$   
	& 3.34     & 3.69     & 9.93
	& \textcolor{blue}{\underline{10.90}}    & 4.20     & 10.63
    & 10.43         & 10.60
    & 2.65     & 4.33  & 8.86  & 10.24    &  9.95
    & \textcolor{red}{\bm{$11.28$}}    \\

    \multirow{6}{*}{}  
	&	$R@10$   
	& 8.36     & 9.17     & 15.22
	& \textcolor{blue}{\underline{18.02}}    & 10.38     & 17.72
    & 16.76        & 17.74
    & 7.79     & 10.67   & 14.43   & 16.95     &  17.14
    & \textcolor{red}{\bm{$18.52$}}   \\

    \midrule

    \multirow{6}{*}{\emph{ML-10M}}  
	&	$N@3$   
	& 3.28    & -     & 45.30
	& 53.47   & 7.78  & 52.16
    & 49.06     & 42.88    
    & 15.23     & 20.27    & 47.12    & 51.80   &  \textcolor{blue}{\underline{53.92}}
    & \textcolor{red}{\bm{$54.75$}}    \\
	
	\multirow{6}{*}{}  
	&	$N@5$   
	& 4.35    & -     & 43.16
	& 51.02   & 10.21  & 49.86
    & 45.17     & 41.24    
    & 17.05     &  22.68   & 44.92   & 49.54    &  \textcolor{blue}{\underline{51.36}}
    & \textcolor{red}{\bm{$52.26$}}    \\

	\multirow{6}{*}{}  
	&	$N@10$   
	& 6.86    & -     & 41.54
	& 47.26   & 14.88  & 46.28
    & 42.33     &  38.65   
    & 19.44    &  25.36    & 42.15   & 46.16   &  \textcolor{blue}{\underline{47.79}}
    & \textcolor{red}{\bm{$48.44$}}    \\
	
	\multirow{6}{*}{}  
	&	$R@3$   
	& 0.12    & -     & 4.67
	& 5.24  & 0.86  & 5.19
    & 4.90     & 3.87    
    & 1.38     &  2.15    & 4.50   & 5.12   &  \textcolor{blue}{\underline{5.54}}
    & \textcolor{red}{\bm{$5.62$}}    \\

    \multirow{6}{*}{}  
	&	$R@5$   
	& 1.25    & -     & 7.82
	& 7.73  & 2.50  & 7.99
    & 7.57     & 5.97    
    & 3.56     &  4.25    & 7.07  & 7.87   &  \textcolor{blue}{\underline{8.10}}
    & \textcolor{red}{\bm{$8.55$}}    \\

    \multirow{6}{*}{}  
	&	$R@10$   
	& 2.36    & -     & 12.56
	& 13.95   & 4.04  & 13.76
    & 12.78     & 10.46    
    & 6.02     & 6.83     & 12.28  & 13.58   &  \textcolor{blue}{\underline{14.26}}
    & \textcolor{red}{\bm{$14.71$}}    \\

	\bottomrule	
				
	\end{tabular}
}
\label{comlete-top-N}
\end{table*}

\begin{table*}[!t]
    \centering
    \caption{Link prediction performance (\%). ($ROC$: \emph{AUC-ROC}, $PR$: \emph{AUC-PR})}
		
    \renewcommand{\arraystretch}{1.2} 
    \setlength\tabcolsep{3pt} %
    \resizebox{\linewidth}{!}{
    \begin{tabular}{c|c|cccccccccccccc}
	\toprule
	
	\textbf{Datasets}
	&    \textbf{Metrics}   
	&    \emph{LINE}    &  \emph{BiNE}           &   \emph{GEBEp} 
	&    \emph{LightGCN}      &  \emph{BiGI}     &   \emph{SHT}     
    &    \emph{HCCF}          &   \emph{AdaGCL}  
    &  \emph{GEL}      &   \emph{STERLING}   &  \emph{DR-GNN}  &    \emph{LightGCL}     
    &   \emph{TransGNN}
    &    \emph{CCBIE}   \\ 
	
	\midrule
				
	\multirow{2}{*}{\emph{Wikipedia}}  
	&	$ROC$   
	& 96.33     & 95.64     & \textcolor{blue}{\underline{96.42}}
	& 95.53    & 95.40     & 94.28
    &  95.37    &  95.06   & 92.84
    &  91.85    &  96.08    & 95.68  & 96.34
    &  \textcolor{red}{\bm{$96.96$}}   \\
	
	\multirow{2}{*}{}  
	&	$PR$   
	& 96.51     &  96.40    & 96.53
	& 96.44    & 96.01     & 94.76
    & 95.66     &  95.33   & 93.80
    &  92.09    & 96.17     & 95.88  & \textcolor{blue}{\underline{96.64}}
    &  \textcolor{red}{\bm{$97.28$}}  \\

	\midrule

    \multirow{2}{*}{\emph{Yelp}}  
	&	$ROC$   
	& 61.06     &  74.31    & 78.62
	& 82.66    &  72.19    & 84.74
    & 82.79    &  80.06   
    & 75.05     & 80.92     & 81.66   & 84.02   &  \textcolor{blue}{\underline{85.17}}
    & \textcolor{red}{\bm{$86.75$}}    \\
	
	\multirow{2}{*}{}  
	&	$PR$   
	& 62.11     & 72.94     & 76.17
	& 83.83    & 70.54    &  86.32
    & 84.16     &  81.34   
    & 72.62     &  79.23    & 80.24   &  84.65   &  \textcolor{blue}{\underline{86.64}}
    & \textcolor{red}{\bm{$87.81$}}   \\
	
    \midrule

    \multirow{2}{*}{\emph{Pinterest}}  
	&	$ROC$   
	& 77.62     &  69.40    & 94.11
	& 94.38    & 80.18     & 95.16
    & 94.02     & 91.38    
    & 82.86     & 84.26     & 92.97   & 94.83  & \textcolor{blue}{\underline{95.29}}
    & \textcolor{red}{\bm{$96.03$}}    \\
	
	\multirow{2}{*}{}  
	&	$PR$   
	& 75.23     &  67.09    & 93.77
	& 93.56    &  78.16    & 94.05
    & 92.25     & 89.26    
    & 80.11     & 82.79     & 91.85  &  93.22   & \textcolor{blue}{\underline{94.43}}
    & \textcolor{red}{\bm{$95.06$}}   \\
	
	\bottomrule	
				
	\end{tabular}
}
\label{LINK-PREDICTION}
\end{table*}

\subsection{Experimental Results}


\textbf{Top-$K$ Recommendation.} Table \ref{comlete-top-N} details the top-$K$ recommendation performance of \emph{CCBIE} and all competitors across three bipartite graphs. The results unequivocally demonstrate that \emph{CCBIE} surpasses all 13 baseline methods across all evaluation metrics on all datasets. We consistently observe that \emph{GNN}-based methods exhibit superior performance compared to metric-based methods in nearly all instances. This highlights the capability of advanced deep learning techniques to autonomously capture complex structural information, leading to the generation of superior-quality embeddings for bipartite graphs. Our \emph{CCBIE} demonstrates substantial superiority over the best competitor, \emph{TransGNN} (global-local learning method), across various graphs, particularly excelling on the dense graph \emph{ML-100K} with a 16.90\% increase in $NDCG@3$. While \emph{CCBIE} exhibits a marginal deficit compared to \emph{TransGNN} in the $RECALL@K$ metric on the sparse graph \emph{DBLP}, with a gap of up to 2\% ($RECALL@5$), it significantly outperforms \emph{TransGNN} in terms of the $NDCG@K$ metric. Concretely, \emph{CCBIE} demonstrates a 10.35\% improvement in $NDCG@3$, 7.19\% in $NDCG@5$, and 4.08\% in $NDCG@10$. This enhancement is even more pronounced compared to other existing global-local methods such as \emph{LightGCL}, \emph{SHT}, and \emph{AdaGCL} and extremely competitive local learning model \emph{LightGCN}. These results affirm the effectiveness of our proposed global-local learning framework in modeling users and items, especially in scenarios with sparse historical interactions.


It is noteworthy that existing model \emph{CEL-NMF} and \emph{STERLING}, which also attempt to incorporate community structure to enhance node embedding quality, exhibit notably poorer performance in experiments compared to other global-local based models, particularly within the dense network \emph{ML-100K}. \emph{CEL-NMF} decomposes the item embedding matrix into a product aligned with a clustering structure, iterating between embedding and cluster optimizations to reconstruct the edge matrix. However, this approach not only risks falling into local optimal solutions but also faces a critical challenge with hard cluster allocation. This method assumes that the embeddings of nodes within the same cluster are identical, significantly compromising the distinguishability between different nodes. \emph{STERLING} also endeavors to capture global synergies using clustering structures, focusing on maximizing the mutual information among co-clusters of users and items. However, we observe that this approach restricts individual users from personalizing different clusters of items, thereby potentially compromising the retention of valuable structural information.

\textbf{Link Prediction.} As demonstrated in the experimental results presented in Table \ref{LINK-PREDICTION}, \emph{CCBIE} consistently outperformed all baseline methods across various datasets and assessment metrics. In particular, on the widely compared dataset \emph{Wikipedia}, \emph{CCBIE} achieves 96.96\% \emph{AUC-ROC} and 97.28\% \emph{AUC-PR}, whereas the top competitor \emph{GEBEp} has 96.42\% \emph{AUC-ROC} and 96.53\% \emph{AUC-PR}, respectively. In summary, \emph{CCBIE}, incorporating clustering constraints, generates higher-quality embeddings that enhance performance in downstream tasks. The experimental results on both tasks provide robust evidence supporting the efficacy of our proposed techniques in both global and local learning contexts, particularly within sparse bipartite graph learning scenarios.

\begin{figure}[htbp]
    \centering
    \begin{minipage}{0.49\linewidth}
	  \centering
	  \includegraphics[width=0.99\linewidth]{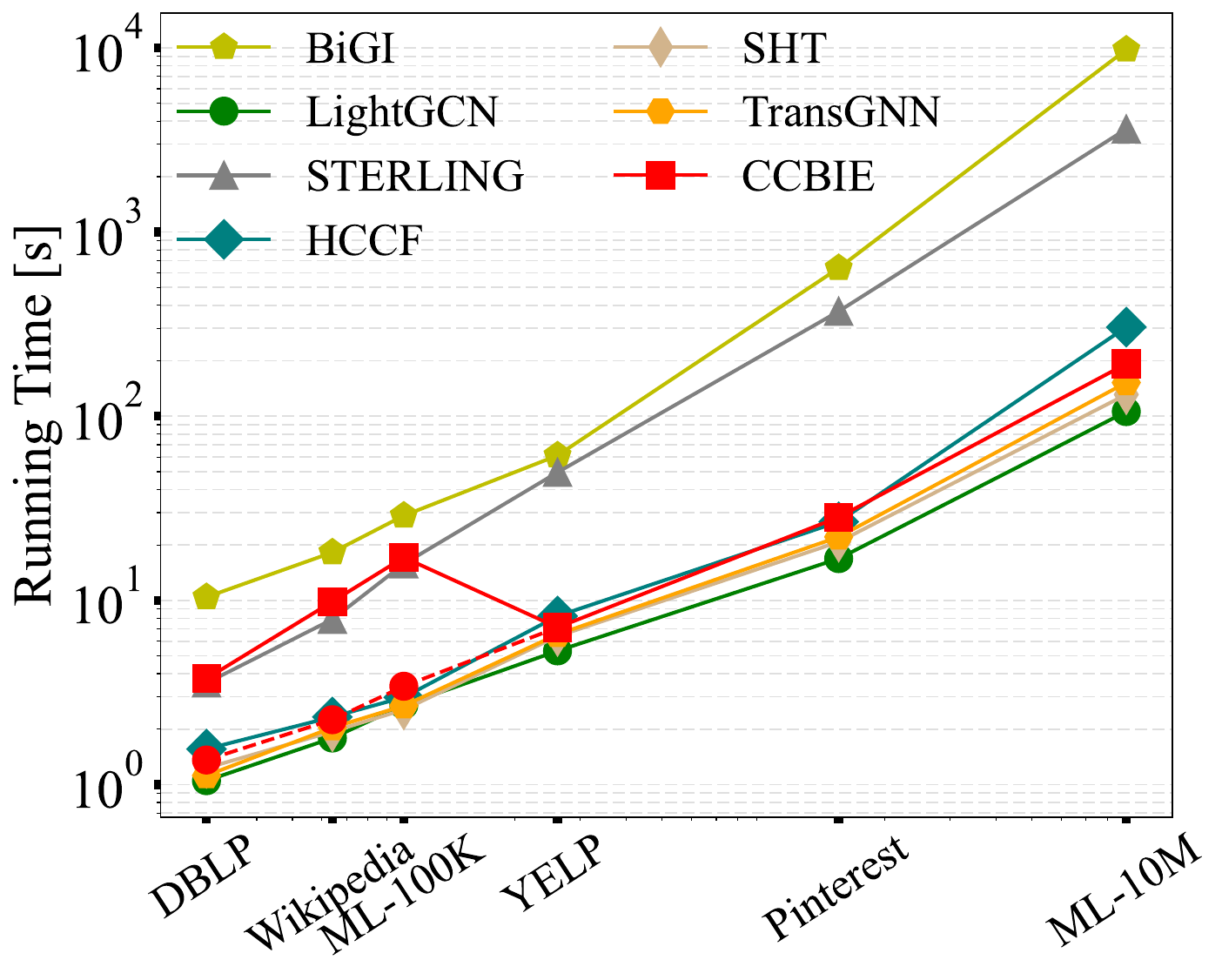}
        \caption{Running time.}
        \label{Running time}
    \end{minipage}
    \begin{minipage}{0.49\linewidth}
	  \centering
	  \includegraphics[width=0.99\linewidth]{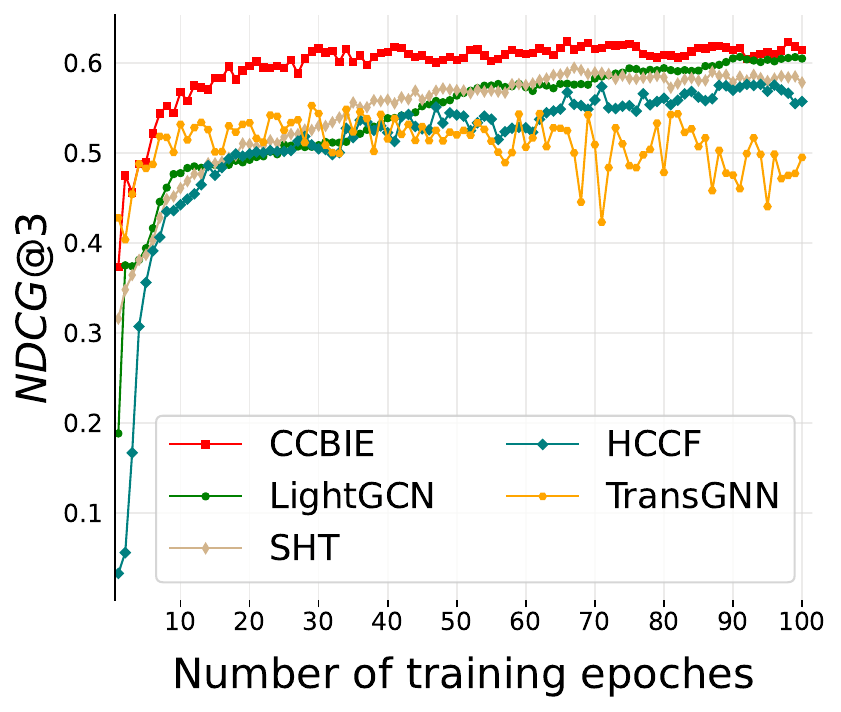}
        \caption{Training curves.}
        \label{training_curves}
    \end{minipage}
\end{figure}

\textbf{Scalability.} The time complexity of \emph{CCBIE} in a mini-batch manner depends solely on hyperparameters, offering increased feasibility and efficiency compared to neighborhood-based \textbf{\emph{MP}} when dealing with a large number of edges. In Fig.\ref{Running time}, we provide the scalability evaluation of \emph{CCBIE} and competitors by varying the number of edges in the input bipartite graph. The reported running times pertain solely to the individual cycles of each model, explicitly excluding the time taken for dataset loading and performance validation. Note that for small-scale networks, we implement an advanced negative sampling strategy designed to enhance the prediction accuracy of the model by ensuring the validity of all negative node pairs generated during sampling, i.e., $\forall u_{i}, \mathcal{N}_{\mathcal{S} } \left ( u_{i} \right )\cap \textbf{E}= \emptyset $. While this procedure does result in increased training time, we believe that the benefits in terms of model performance justify the additional computational cost. We also present the runtimes associated with the plain sampling approach, as indicated by the red dashed line. As depicted in Fig.\ref{Running time}, across all six datasets utilized for comparative analysis, the performance of \emph{CCBIE} consistently lags behind that of the leading \emph{GNN}-based competitors, albeit within the same order of magnitude. However, as the network scale expands, the advantage in efficiency of \emph{CCBIE} become increasingly evident, owing to its reduced dependence on full-edge training on large graphs.

Furthermore, Fig.\ref{training_curves} displays the training curves of competitors on \emph{ML-100K}. The findings reveal that \emph{CCBIE} converges faster and exhibits superior performance compared to current SOTA models, underscoring the robust generalization ability facilitated by cluster-based \textbf{\emph{MP}}. Specifically, \emph{CCBIE} demonstrates optimal training performance with minimal training batches, whereas \emph{LightGCN}, for instance, necessitates nearly 100 batches to achieve comparable results. Interestingly, \emph{TransGNN} demonstrates a convergence rate comparable to that of \emph{CCBIE}, indicating that a broader acceptance domain may facilitate convergence. However, \emph{TransGNN}'s performance is markedly unstable, likely due to its reliance on transformers for global structural information, which may cause excessive focus on long-range dependencies. Conversely, \emph{CCBIE} captures the network's global structure at the meso-scale through clustering, enabling a more refined integration of contextual topological information.

\textbf{Ablation Study.} To rigorously evaluate the effectiveness of our proposed global and local learning components, we meticulously designed four variants of \emph{CCBIE}. The variants are described in detail as follows: (1) $\emph{CCBIE}_{E}$: a variant without implicit relationships, i.e, $\textbf{Y}= \widetilde{\textbf{Y}}$; (2) $\emph{CCBIE}_{I}$: a variant without explicit relationships, i.e, $\textbf{Y}= \widehat{\textbf{Y}}$; (3) $\emph{CCBIE}_{M}$: a variant that omits the augmentation of $\textbf{T}^{M}$ and $\textbf{P}^{M}$ using multilayer perceptron, i.e, $\textbf{T}^{M} = \textbf{T}$ and $\textbf{P}^{M} = \textbf{P}$; (4) $\emph{CCBIE}_{S}$: a variant that computes the cluster affiliation of items using softmax function instead of Eq.\ref{uni-max}, i.e., 
\begin{equation}
    \textbf{T}^{V}_{ij} = \frac{exp\left ( \textbf{T}^{O}_{ij} \right ) }{ {\textstyle \sum_{l=1}^{K_{q}}} exp\left ( \textbf{T}^{O}_{il} \right )  }\
    \label{softmax}
\end{equation}

\begin{figure}[htbp]
    \centering
    \includegraphics[width=0.90\linewidth]{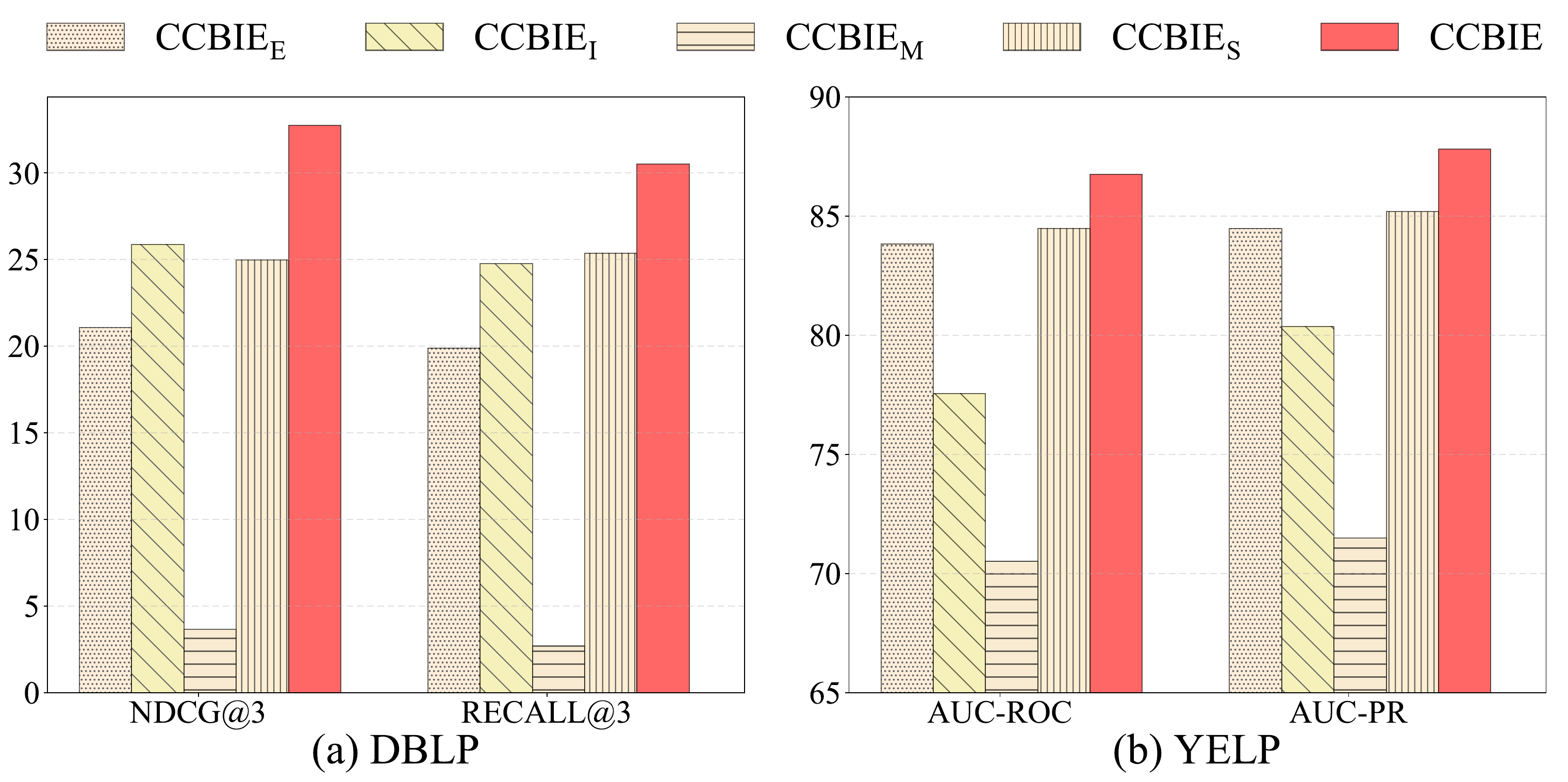}
    \caption{Ablation study of \emph{CCBIE} in terms of accuracy.}
    \label{VAR-MODELS}
\end{figure}

From the findings presented in Fig.\ref{VAR-MODELS}, it is apparent that \emph{CCBIE} demonstrates superior performance relative to models such as \emph{BiGI} across most scenarios solely through implicit relation learning ($\emph{CCBIE}_{I}$). This underscores the effectiveness of higher-order structural learning within bipartite graphs. Moreover, the model's competitive strength when relying solely on explicit and implicit relation learning is not particularly pronounced, but significant enhancement is achieved with the adoption of straightforward joint training techniques. This observation indicates that the local-global learning paradigm has the capability to capture more meaningful structural information, thereby contributing to higher-quality node embeddings. An important finding of this study is that the $\emph{CCBIE}_{I}$ model exhibits a substantial advantage over the $\emph{CCBIE}_{E}$ model in the top-$N$ recommendation task, whereas the reverse holds true for the link prediction task. This aligns with the intuition that explicit relations are adept at capturing similarities between pairs of nodes, which enhances the accuracy of local structural representation and is thus more effective for link prediction. Conversely, implicit relations capture node similarities on a mesoscopic scale, offering a more holistic depiction of contextual node information, which proves advantageous for personalized ranking tasks. This implies that the value of parameter $\alpha$ can be adjusted depending on the downstream task. For tasks such as link prediction, reducing $\alpha$ directs \emph{CCBIE} to prioritize local structure. Conversely, in recommender systems, a higher value of $\alpha$ can be specified.

Furthermore, comparing with $\emph{CCBIE}_{M}$ indicates that employing \emph{MLP} to augment node clustering information can notably enhance the performance of \emph{BGE}. This highlights the limitation of relying on a single similarity metric to fully capture the intricate relationships between nodes and clusters. Finally, we observe a notable performance degradation of $\emph{CCBIE}_{S}$ compared to \emph{CCBIE}, with improvements of up to 31.01\% in recommendation and 2.67\% in link prediction. This motivates us to pursue a more restrained computational method for determining the cluster affiliation of nodes.

\begin{figure}[htbp]
    \centering
    \includegraphics[width=0.99\linewidth]{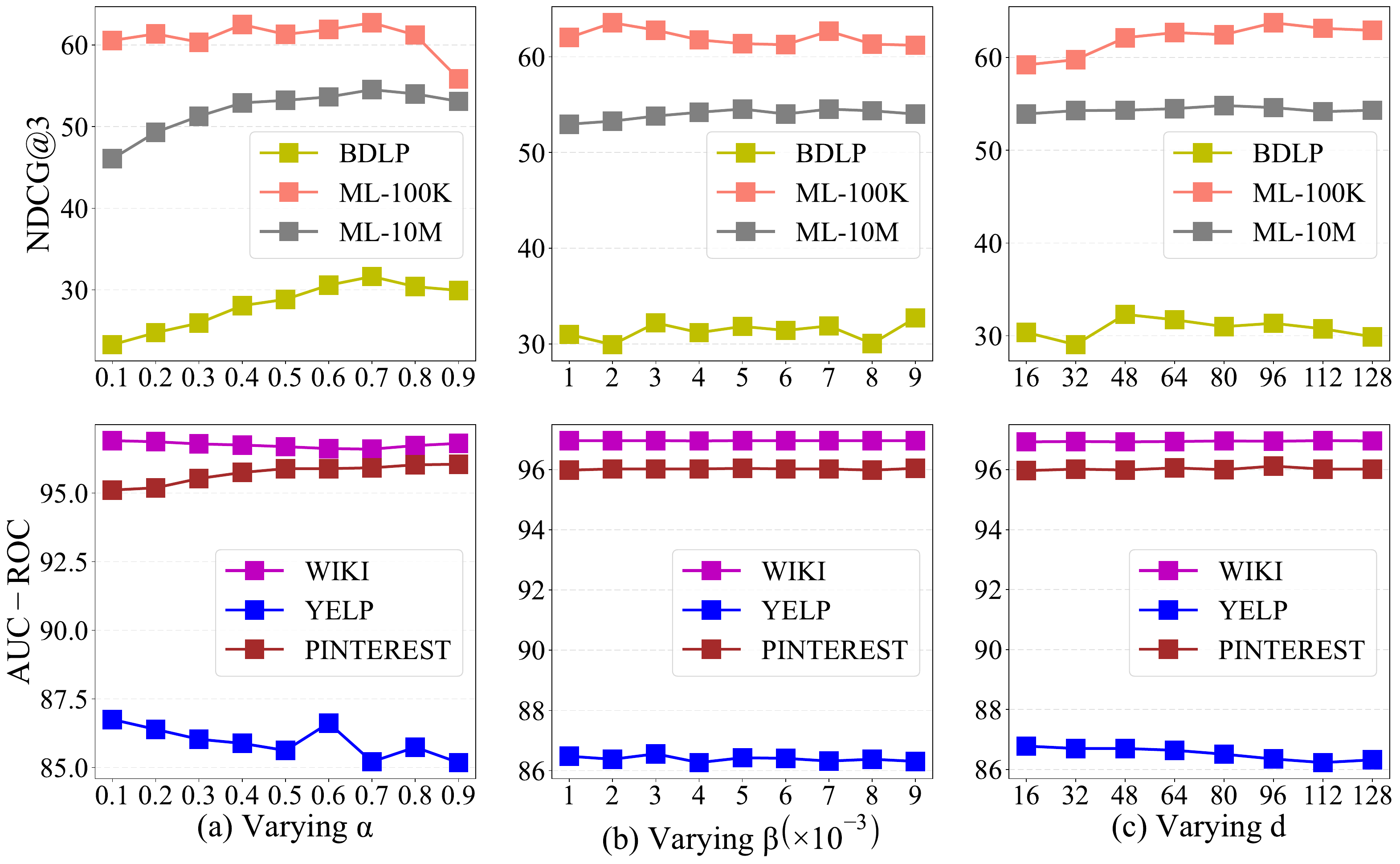}
    \caption{Predictive performance when varying $\alpha$, $\beta$ and $d$.}
    \label{CHANGE_PARAMS}
\end{figure}

\textbf{Hyper-parameter sensitivity.} We conducted additional investigation into the impact of varying \emph{CCBIE} parameters, specifically parameter $\alpha$ in Eq.\ref{relations}, which governs the relative importance of explicit/implicit relationships; parameter $\beta$ in Eq.\ref{cluster-sep}, which regulates the degree of cluster separation; and parameter $d$, dictating the expressiveness of \emph{CCBIE}. We conduct evaluations on the six datasets for the top-$N$ recommendation and link prediction task, and document the results of varying parameters $\alpha$, $\beta$ and $d$ in Fig.\ref{CHANGE_PARAMS}. Based on the experimental findings, it is evident that:


\textbf{Sensitivity of $\alpha$.} As $\alpha$ increases, \emph{CCBIE} prioritizes local structural information, leading to the degeneration of \emph{CCBIE} into $\emph{CCBIE}_{E}$ when $\alpha$ reaches 1. Conversely, decreasing $\alpha$ to 0 emphasizes the significance of global cluster synergies in hypothesizing relationships between node pairs, causing \emph{CCBIE} to degenerate into $\emph{CCBIE}_{I}$. Based on the experimental findings, it is evident that a larger $\alpha$ value contributes significantly to enhancing the recommendation performance of \emph{CCBIE}. Nevertheless, an excessively large $\alpha$ can cause \emph{CCBIE} to excessively prioritize global information, thereby impairing its overall performance. In our experiments, we empirically set $\alpha = 0.7$ for the recommender system task.

In the context of link prediction, the values of parameter $\alpha$ are intricately dependent on the network's structure, adding complexity to the analysis: optimal $\alpha$-value of 0.1 is observed for the sparser bipartite graphs \emph{Wikipedia} and \emph{Yelp}, whereas in the denser bipartite graph \emph{Pinterest}, setting $\alpha=0.6$ yields the best prediction performance. We hypothesize that this phenomenon stems from sparse graphs exhibiting weaker clustering phenomena, thus offering less meso-scale structural information for node similarity. Consequently, implicit relationships play a diminished role, necessitating greater attention to explicit relationships. Conversely, dense graphs furnish ample higher-order structural insights for effective modeling, where explicit relationships alone may not sufficiently differentiate node pairs, warranting a heightened emphasis on explicit relationships.

\textbf{Sensitivity of $\beta$.} A higher value of $\beta$ indicates a more distinct clustering structure, enhancing the differentiation between nodes. The experimental results reveal a heightened sensitivity to parameter $\beta$ for the recommendation task on \emph{DBLP} and \emph{ML-100K}, while its impact is minimal in other contexts. A conspicuous finding is that the values of parameter $\beta$ produce markedly different trends on the $NDCG@3$ metric for the two networks: the same parameter that improves prediction performance for \emph{DBLP} results in decreased accuracy for \emph{ML-100K}. This disparity is intriguing, given that \emph{DBLP} and \emph{ML-100K} represent a sparse and dense bipartite graph, respectively, suggesting a potential regularity in how clustering structures influence node embedding quality across varying network characteristics. Investigating this systematic pattern will be the focus of our subsequent research phase. For a balanced assessment across various networks, we set $\beta=0.005$ by default.

\textbf{Sensitivity of $d$.} The embedding dimension significantly influences the expressiveness of \emph{CCBIE}, with higher values of $d$ generally resulting in improved performance across various metrics. Nonetheless, our observations reveal a nuanced trend: as $d$ increases, performance initially improves up to a critical point, beyond which it begins to deteriorate. This phenomenon can be attributed to either incomplete training or an overly dispersed distribution of nodes within the high-dimensional embedding space. Throughout our experiments, we maintained a fixed embedding dimension of $d = 64$.

\section{Conclusion}

In this paper, we present a novel bipartite graph embedding model called \emph{CCBIE} that incorporates clustering constraints. \emph{CCBIE} effectively addresses the challenge of alleviating cold users and items by leveraging integrated global and local learning mechanisms to produce high-quality node embeddings. Numerous experiments validate the superior performance of \emph{CCBIE} in terms of both embedding quality and model scalability. 

\bibliographystyle{IEEEtran}
\bibliography{sn-bibliography}

\vfill

\end{document}